\documentclass[11pt,twoside]{article}

\usepackage{asp2006}
\usepackage{epsf}
\usepackage{psfig}
\usepackage{lscape}

\begin{document}
\title{Bohdan's Impact on Our Understanding of Gamma-ray Bursts
}   
\author{Tsvi Piran }   
\affil{Racah Institute for Physics, The Hebrew University, Jerusalem, Israel 91904}    

\begin{abstract} 
Bohdan Paczy{\'n}ski was one of the pioneers of the cosmological GRB
model. His ideas on how GRBs operate and what are their progenitors
have dominated the field of GRBs in the hectic nineties during which
the distances and the origin of GRBs were revealed. I discuss here
Bohdan's contributions in some historical perspective.

\end{abstract}


\section{Prologue}   

I first met Bohdan in the summer of 1977 at the IOA. I don't
remember the exact date, but I remember the time. It was between 3am
and 4am. Several of us were trying to open the last remaining bottle
of wine at the traditional IOA summer (Argentinean) barbecue.
Something did not work out. Bohdan approached us and succinctly
pointed out that the process will be much easier if we remove first
the plastic warp that covered the cork. We did. It worked. This was
a simple demonstration of Bohdan's ability to see the obvious simple
fact that everyone else around stared at - but no one sees.

Our next meeting took place several years later at Princeton. Bohdan
and his family moved to the Institute housing, where I lived as
well. I have spent many pleasant evenings visiting them. We enjoyed
many discussions. When we did not discuss life in America versus
life in Europe those conversations turned naturally to Gamma-Ray
Bursts (GRBs) - a subject that fascinated both of us.

\section{Introduction - GRBs as we understand them today}

I begin with a brief description of the present picture of GRBs and
their models. This section is not intended to be a comprehensive
review (see Piran 1999; M{\'e}sz{\'a}ros 2002; Piran 2005;
M{\'e}sz{\'a}ros 2006; Woosley \& Bloom 2006; Zhang, 2007 for recent
reviews and Fishman \& Meegan, 1995; Piran 1995 for a historical
perspective). My objective is to outline our current understanding
so that Bohdan's contributions and impact can be put into the right
perspective and be appreciated.

GRBs are short and intense ($\sim 10^{-7}-10^{-5}$ergs/cm$^2$ )
bursts of soft (10 keV - 2 MeV) gamma rays coming from random
directions in the sky. GRBs were discovered accidentally in the late
sixties by the Vela satellites - defense satellites that were sent
to monitor the outer space treaty. The discovery was reported only
in 1973 (Klebesadel et al.). In the early nineties  the BATSE
detector onboard of the COMPTON-GRO satellite indicated that GRBs
are cosmological (Meegan 1992, Mao \& Paczy{\'n}ski 1992, Piran
1992). Towards the late nineties the BeppoSAX satellite discovered
X-ray afterglows (Costa et al., 1997). Optical (van Paradijs et al.,
1997) and Radio (Frail and et al., 1997) afterglows followed
shortly. Direct redshift measurement of the optical afterglow of GRB
970508 (Matzger et al., 1997) established the cosmological origin of
GRBs. Radio scintillation (Goodman, 1997) and later VLBI
observations (Taylor et al., 2004) confirmed the relativistic bulk
motion, predicted earlier by the fireball model. In the following
years afterglow observations revealed a wealth of information on
host galaxies (Djorgovski et al. 2003) and the positions of GRBs
within host galaxies, which was always within star forming regions
(Paczy{\'n}ski 1998, Fruchter, 2006).

GRB belong to two groups (Kouveliotou et al. 1993) according to
their durations $T_{90}$:  Long GRBs (with $T_{90}>2$sec) and short
(with $T_{90}<2$). Short bursts are also typically harder and hence
they are called at times short hard bursts (SHBs). The association
of several long GRBs with supernovae (Galama et al. 1998; Bloom et
al. 1999; Stanek et al. 2003; Hjorth,  et al. 2003) confirmed the
Collapsar model (Woosley 1993; Paczy{\'n}ski 1998; McFadyen \&
Woosley, 1998) according to which (long duration) GRBs arise during
the collapse of massive stars. The origin of short GRBs is less
certain. Unlike long GRBs, short ones do not arise in star forming
regions. They arise in both elliptical and spiral galaxies (see e.g.
Nakar 2007 for a review). The redshift distributions shows that the
observed short bursts are much nearer than the observed long ones
(Guetta \& Piran, 2006; Nakar et al., 2006). These observations are
consistent with earlier suggestions that short GRBs arise from
neutron star mergers (Paczy{\'n}ski, 1986; Eichler et al., 1999).
Within this model the delay corresponds to the gravitational
radiation spiral in time. The delay is crucial also in positioning
the progenitors away from their birth places, possible even outside
the original galaxy.

The  internal-external shocks model (also called the fireball model)
(Goodman 1986; Paczy{\'n}ski, 1986; Shemi \& Piran, 1990;
M{\'e}sz{\'a}ros \& Rees 1992, Rees \& M{\'e}sz{\'a}ros, 1994,
Narayan, Paczy{\'n}ski \& Piran, 1992; Sari \& Piran 1997, Wijers,
Rees \& M{\'e}sz{\'a}ros, 1997) is depicted in Fig. 1. It suggests
that GRBs involve ultrarelativisitic (baryonic or Poynting flux)
jets that emerge from compact objects, most likely black holes.
Internal collisionless shocks or instabilities  within the
relativistic outflow, that take place at $10^{14}-10^{16}$cm from
the central engine, accelerate particles to relativistic velocities
and generate strong magnetic field. Synchrotron or SSC emission of
these particles produce the prompt emission. The outflow interacts
later with the surrounding medium producing a relativistic blast
wave -  a relativistic analogue of a Supernova remnant. This long
lasting blast wave accelerates particles and generates magnetic
fields that produce the multiwavelength afterglow. The predictions
(Sari, Narayan \& Piran, 1998; Sari, Halpern \& Piran, 1999) of this
"standard model"  agreed nicely the observed light curves and
spectra of GRB afterglows (Wijers \& Galama, 1999; Kulkarni et al.,
1999; Panaitescu \& Kumar 2002; Yost S.A., et al., 2003). Recent
{\it Swift} observations indicate that the picture is somewhat more
complicated than what is described here.  Still for the purpose of
this review this description is sufficient.

\begin{figure}[!ht]

\plotfiddle{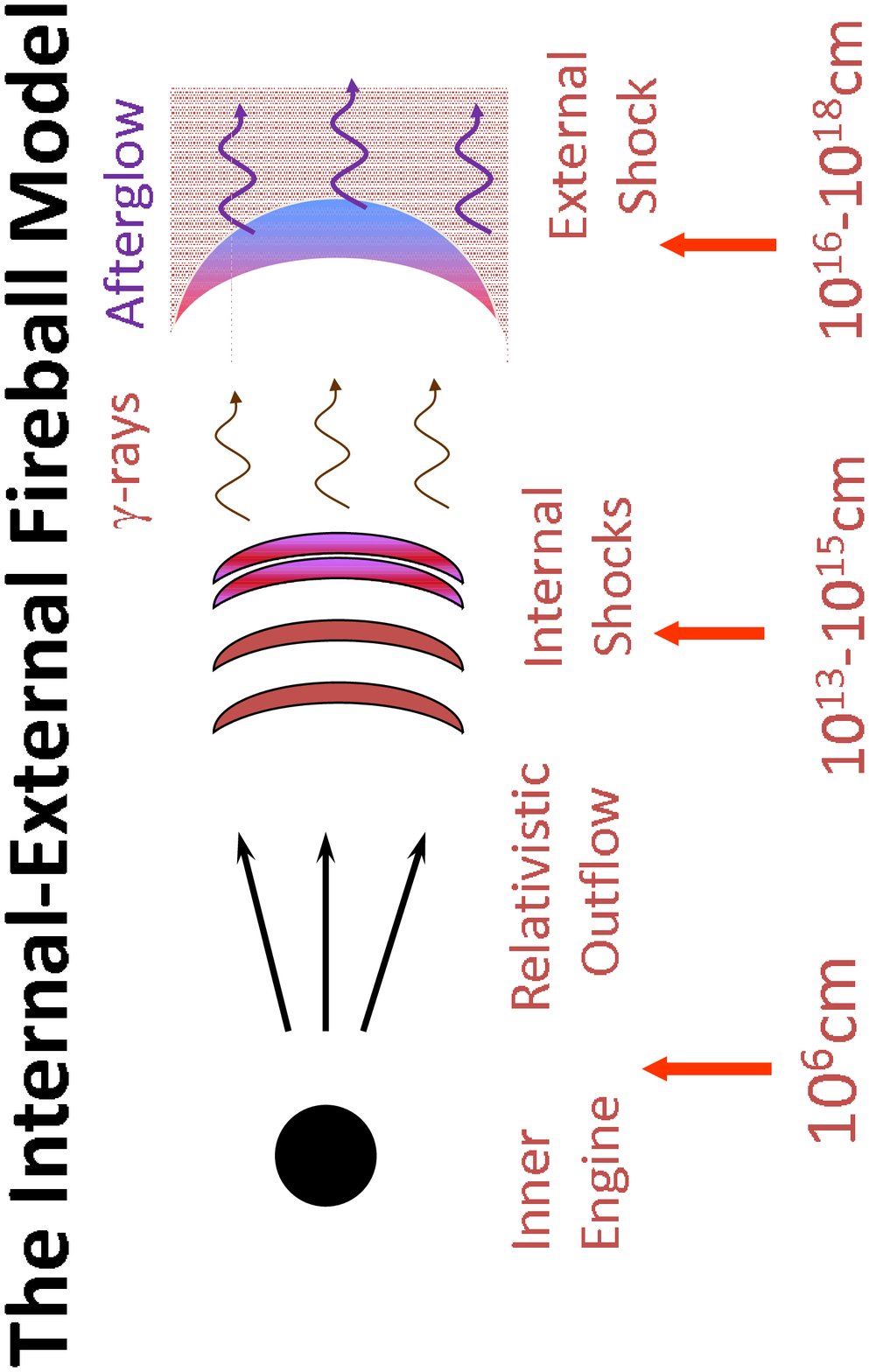}{10cm}{-90}{50}{50}{-200}{250}
\caption{The standard internal-external shocks model of GRBs (see
e.g. Piran 1999; M{\'e}sz{\'a}ros 2002; Zhang and  M{\'e}sz{\'a}ros
2004; Piran 2005; M{\'e}sz{\'a}ros 2006 for reviews). 1.  Central
engine and acceleration of the flow. 2.  Coasting phase; terminal
Lorentz factor $ \Gamma> 100$. 3.  Internal shocks; prompt gamma-ray
emission . 4.  The interaction of the outflow with the surrounding
matter produces the external shocks system with a blast wave
propagating outwards producing the afterglow. }
\end{figure}

\section{Prehistory - GRBs before BATSE}

The history of GRBs should be divided to two period. Pre and post
BATSE. At the pre-BATSE era (up to the late eighties and very early
nineties) a consensus formed that GRBs originate in magnetic
instabilities on galactic neutron stars. What was considered as a
strong evidence for this model was the detection of 20 and 40 keV
lines which were interpreted as cyclotron lines of a $\sim 10^{12}$
Gauss field - a magnetic field expected in neutron stars.

Bohdan was skeptical. He never believed the observed 20 and 40 keV
lines e.g.(see Narayan, Paczy{\'n}ski \& Piran,  1992) - insisting
that there was never the case that the {\it same} line was observed
{\it simultaneously} from the {\it same} burst by {\it two
different} detectors. With no lines there was no compelling case for
galactic neutron stars and the road to cosmological sources was
open. If the bursts are cosmological they release much more energy
$\sim 10^{51}$ergs or more. A basic problem with any cosmological
model was that GRBs have non-thermal spectra indicating that they
are optically thin. However, the temporal variability (less than a
fraction of a second) indicated that the emission region is small
(less than $10^9$cm). If $10^{51}$ergs of soft gamma-rays (with
numerous photons above 500 keV) are released from such a source
there would be copious pair production that will lead to an
optically thick source. This is the so called "Compactness problem".

\begin{figure}[!ht]
\plotfiddle{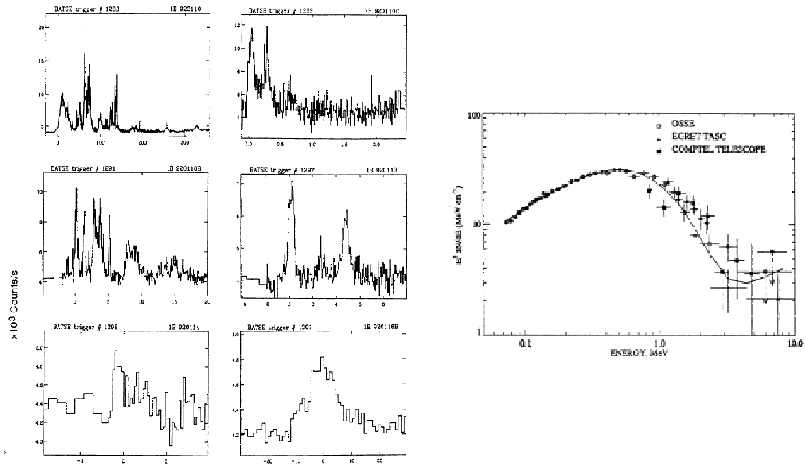}{10cm}{-90}{50}{50}{-200}{250}
\caption{The ``Compactness problem". Left: light curves for
different bursts. Variability on very short time scales is clearly
seen. Right: The spectrum of GRB 910601. The nonthermal spectrum is
clearly seen. If $10^{51}$ergs of radiation with many photons above
$m_e c^2$ (as seen in this spectrum) are released within a small
region (as indicated by the temporal variability) numerous pairs
will form and the radiation-pairs plasma will be optically thin.}
\end{figure}

The breakthrough concerning the resolution of the "Compactness
problem"  came in  1986 when two back to back papers where publisehd
in the Astrophysical Journal Letters by Jeremy Goodman (1986) and
Bohdan Paczy{\'n}ski (1986). Both papers describe an idealized
problem  in which a hot radiation, a fireball, with $T_0 \gg m_e
c^2$ is released within a small region. They argues that the
radiation will form a radiation-pair plasma fluid that will expand
rapidly. As the fireball accelerates it cools (in its local frame).
When the local temperature drops below $m_e c^2$ the pair begin to
annihilate. There are enough pairs to keep the fireball optically
thick until the local temperature drops to $\sim 20$keV when the
fireball become optically thin. By this time the fireball has
reached a relativistic motion with $\Gamma \sim T_0/(20 {\rm keV})$.
The photons escape freely now. Because of the relativistic motion
the escaping photons will be seen by an observer at rest with
approximately the original temperature $T_0$. Jeremy solved the
problem in the impulse approximation, considering the case of an
``explosion". Bohdan, on the other hand, considered a steady state
approximation in which the evolution was approximated as a steady
wind. The results were more or less similar. The papers were the
first to suggest that {\it GRBs involve relativistic motion}. A
concept that was verified more than ten years later first using
radio scintillations (Goodman, 1997) and later with direct VLBI
measurements (Taylor et al.,  2004). While the two papers are
closely related, they are not similar. Goodman focuses on the
physics and is very concerned with the fact that at the end the
resulting photons have a thermal like spectrum. Bohdan, on the other
hand, is not so worried about this issue and he uses this model to
put forwards his idea that GRBs are in fact cosmological. He even
goes on to suggest, in the discussion section of his work, that some
reoccurring sources arise because of gravitational lensing of
cosmological events.

At that time we spent many hours discussing what could drive such
events. Our discussions focused on neutron star mergers. Objects
that at the time became popular among relativists as it was realized
that they are prime candidates for detection as sources of
gravitational radiation. Our discussions evolved around the final
stages in which tidal interaction between the two stars will tear
them apart, and Bohdan borrowed here a lot from his work on
binaries. Still Bohdan was reluctant to publish a paper  with a
model describing mergers as GRB sources. My guess was that what he
really cared most at the time was to "win the cosmological war" on
the distances to the bursts. He was probably worried that linking
the cosmological idea to the very speculative merger model (Neutron
star mergers were considered as somewhat esoteric events in those
days), might weaken his cosmological case. Bohdan speculated on this
merger possibility very briefly in two or three sentences in his
1986 paper and returned to it (actually to a variant - black hole
neutron star merger (Paczy{\'n}ski, 1991)) only in the early
nineties. I went on and published these ideas, with a detailed model
on how a merger can produce a GRB, in 1989 (Eichler, Livio, Piran \&
Schramm 1989). We wrote our joint paper on this subject only in 1992
(Narayan, Paczy{\'n}ski \& Piran, 1992). I will return to this paper
later in this talk.

\section{BATSE, the Cosmological-Galactic War and the Great Debate}

The Burst and Transient Source Experiment, BATSE, on board of the
Compton-GRO satellite, was expected to demonstrate that GRBs are
galactic. BATSE's spectrograph was supposed to detect the lines and
its ability to localize the bursts within a few degrees was
sufficient to demonstrate that the bursts are within the galactic
disk. It is difficult to imagine the shock   when BATSE's first
results were announced in the fall of 1991: { No lines, Isotropic
distribution with no dipole moment (corresponding to our position
relative to the galactic center),  and last but not least a paucity
of weak bursts} (Meegan et al., 1992). The results were accepted
with disbelief and at first were sort of ignored. Most of the talks,
at least most of the theoretical talks, given at the first
Huntsville symposium, that took place in October 1991, dealt with
various aspects of the galactic neutron star model. A whole session
was devoted to magnetospheric instabilities on neutron stars. One
speaker, J. P. Lasota, withdrew his talk on a galactic model and
stated that it became irrelevant in view of the new data. Bohdan
(Paczy{\'n}ski, 1992) and me (Piran, Narayan \& Shemi 1992) were the
only speakers that dealt with cosmological models in that meeting.

Both Bohdan (Mau \& Paczy{\'n}ski, 1992) and me (Piran 1992) rushed
to publish letters on the essential cosmological interpretation of
BATSE's result. We stressed that the combination of an isotropic
distribution, that at that stage was demonstrated by upper limits on
the dipole moment of the distribution (see e.g. Fishman et al.1994;
Hakkila et al.1994) and the paucity of weak bursts, which at that
stage was not demonstrated by a full $ \log N- \log S$ distribution
(see Fig. 3) but by the very low $\langle V/V_{max} \rangle =0.32$
value (see e.g Fishman et al.1994; Hakkila et al.1994) forced a
cosmological distribution. The argument was surprisingly simple. If
the sources are galactic their angular distribution must show a
significant dipole moment, unless (i) Their radial distribution is
extremely local (and we don't see the galactic structure) or (ii)
The radial distances are very large so that the typical distance
compared to our distance from the galactic center is large.
Possibility (i) is ruled out by the $\langle V/V_{max} \rangle$
value. A small industry begum by proponents of the galactic models
that  threw GRB sources to larger and larger distances in the
galactic halo (e.g. Podsiadlowski et al.1995). As the limits on the
dipole moment become tighter and tighter the average distance needed
for the sources continuously increased. I vividly remember Bohdan
joking that with a typical distance of 200kpc one might call the
model cosmological... Additionally, as the distances increased so
did the energy budget until even the galactic models suffered from
the Compactness problem (Shemi \& Piran, 1993).

\begin{figure}[!ht]
\plotfiddle{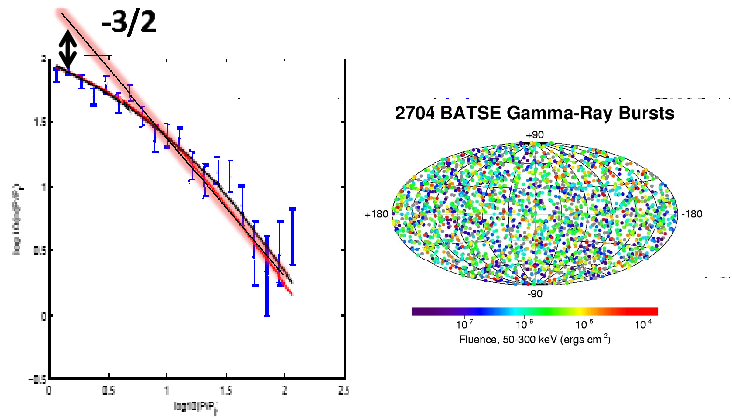}{10cm}{-90}{50}{50}{-200}{250}
\caption{Right: The distribution of BATSE bursts on the sky, in galactic
coordinates. The isotropy is
self evident now. Left: the $\log(N)-\log(S)$ plot. The number of bursts
with $N$ with a peak flux
larger than $S$. In a  homogenous distribution we should see a straight
line with a slope of $-3/2$.
}
\end{figure}

Bohdan was very excited during this period. He enjoyed the lively
(at times too lively) debate that went on. He told me that in order
to prevent misunderstanding or misrepresentations of his ideas he
was signing, at that time,  all GRB referee reports that we wrote.
He was most amused when Nature published a galactic origin paper
(Lingenfelter \& Higdon 1992) and refused to publish his rebuttal
because ``our readers will be confused if we published two
contradicting opinions". Both of us had a great time making fun of
some of the models. Surprisingly the fight went on for quite some
time. At the second Huntsville meeting that took place at the fall
of 1993 more than half of the audience still believed in galactic
origin and this situation persisted even in the ``Great Debate" that
took place in 1995.

The debate culminated in the ``Great Debate" that too place in April
1995 at the Smithsonian in Washington, D.C., between Bohdan
(Paczy{\'n}ski, 1995) and Don Lamb (Lamb 1995). Bohdan's abstract
presents his argument best: {\it ``The positions of over 1000
gamma-ray bursts detected with the BATSE experiment on board of the
Compton Gamma Ray Observatory are uniformly and randomly distributed
in the sky, with no significant concentration to the galactic plane
or to the galactic center. The strong gamma-ray bursts have an
intensity distribution consistent with a number density independent
of distance in Euclidean space. Weak gamma-ray bursts are relatively
rare, indicating that either their number density is reduced at
large distances or that the space in which they are distributed is
non-Euclidean. In other words, we appear to be at the center of a
spherical and bounded distribution of bursters. This is consistent
with the distribution of all objects that are known to be at
cosmological distances (like galaxies and quasars), but inconsistent
with the distribution of any objects which are known to be in our
galaxy (like stars and globular clusters). If the bursters are at
cosmological distances then the weakest bursts should be redshifted,
i.e., on average their durations should be longer and their spectra
should be softer than the corresponding quantities for the strong
bursts. There is some evidence for both effects in the BATSE data.
At this time, the cosmological distance scale is strongly favored
over the galactic one, but is not proven. A definite proof (or
dis-proof) could be provided with the results of a search for very
weak bursts in the Andromeda galaxy (M31) with an instrument ~10
times more sensitive than BATSE. If the bursters are indeed at
cosmological distances then they are the most luminous sources of
electromagnetic radiation known in the universe. At this time we
have no clue as to their nature, even though well over a hundred
suggestions were published in the scientific journals. An experiment
providing ~1 arc second positions would greatly improve the
likelihood that counterparts of gamma-ray bursters are finally
found. A new interplanetary network would offer the best
opportunity."}

It was the 75th anniversary of the Curtis and Shapley 'Great Debate'
(where the size of the Universe was contested just three years
before Edwin Hubble made his seminal discovery about its expansion
and thus the size of the Universe). The parallel of the
Curtis/Shapley and GRB debates could not be more poignant: just as
few changed their minds about the size of the Universe after the
1920 debate as have changed their mind about the distances of GRBs
after this one. The audience was split roughly 50\% 50\% at the end
of this discussion. I must say that  when I heard the results of the
vote at this debate I was shocked. To me at that time the question
what are the distances of GRBs was not an open question (Piran
1995).

Bohdan was very happy about this work. In a report of his
achievements in a NASA grant that he posts (in a very unusual
manner) on astro-ph (Paczy{\'n}ski, 1996) he writes: {\it ``The
research project: 'Models and Scenarios for Gamma-Ray Bursts'
resulted in a total of 20 published research papers. The central
issue was the distance scale to gamma-ray bursters, the issue
brought up by the remarkable discovery of the distribution
properties of gamma-ray bursts by the BATSE on Compton GRO. The last
paper on the reference list is the PI's contribution to the debate
on the distance scale to gamma-ray bursts held in Washington DC on
22 Apr. 22 1995. When this project got started, right after the
announcement of the BATSE results at the conference in Annapolis in
the fall of 1991, only a small fraction of astrophysicists seriously
considered the possibility that gamma-ray bursts are at cosmological
distances. By now the cosmological distance scale has become a
majority view, to a large extent because of the publications listed
in this final report. PI considers this to be the most important and
lasting result from the research supported by this grant."}

The debate ended  with the measurement of the redshift of the
optical afterglow of GRB 970508 (Metzger et al., 1997). It is
amusing to recall that even after this observation   ``conspiracy
theories" suggesting  coincidence between this GRB and the observed
afterglow was put forwards for a while.

\section{Building a theoretical model - how GRBs work}

In 1986 Bohdan contributed to the early ideas on a thermal fireball.
However, it was clear that this idea is rather preliminary. The
resulting emission from this simple fireball has a quasi-thermal
spectrum, quite unlike what is observed. Additionally,it was clear
that this picture of a pure radiation (and pairs) fireball was too
idealized.

\subsection{From Radiation and  Baryons to Ultra relativistic Baryonic outflow}
The next question was what would be the effects of baryons on such a
pure radiation fireball. Here, once more we have worked in parallel
on the same problem and the results were published more or less
simultaneously (Abramowicz, Novikov \& Paczy{\'n}ski, 1990; Shemi \&
Piran, 1990). Both works have shown that the addition of even a
small baryonic load would change drastically the outcome. The
baryons will be dragged along and accelerated by the radiation
field. If the total baryonic load is not too small eventually all
the initial thermal energy will be transferred to the kinetic energy
of the baryons. If the baryonic load is  not too large, that is
$m<E_0 /c^2$ the final outcome will be a relativistic baryonic
outflow with $\Gamma = E_0 /M$ ($E_0$ being the total energy and $M$
the baryonic mass). This was a step in the right direction but it
was not good enough. Relativistic baryons can serve as a source of
cosmic rays. But what we have to find a way to convert their kinetic
bulk energy back to radiation.

\subsection{Astro-ph/9204001}

In the spring of 1992 Bohdan gave a colloquium, on the cosmological
origin of GRBs at Harvard. I was amused by the fact that talk was on
March 5th - the anniversary of the famous 1979 March 5th GRB. This
was was the only well localized GRB at that time, and it was in the
LMC - giving credence to the Galactic origin. By now we know that it
was not a regular GRB but a soft Gamma repeater (SGR) but this is
another story.  Bohdan stayed for a whole week during which we
wrote, together with Ramesh Narayan, {\it ``Gamma-Ray Bursts as the
death throes of massive binary stars"} (Narayan, Paczy{\'n}ski \&
Piran, 1992- NPP92).

This paper, astro-ph/9204001, is the first astro-ph. It is a well
known, highly cited(more than 400 citations) paper. Most remember it
because of the serious attempt to build a consistent merger model
for GRBs. However, in addition to discussing neutron star mergers
that it contained several new ideas that helped shape our
theoretical understanding  of GRBs:
\begin{itemize}
\item Internal shocks
\item GRBs are powered by accretion onto a newborn black holes.
\item The duration of the GRB is determined by the activity of the inner engine
(accretion onto
the black hole) while the duration of the pulses is determined by the dynamical
time scale
of the source
\item GRBs should be followed by a long lasting Afterglow
\item Binary neutron stars has a long spiraling-in phase before they merge.
As they receive a large kick when the binary is born they can escape
during this phase from their host galaxy.
\item $10^{15}$Gauss magnetic fields may exist at GRBs' inner engines.
\end{itemize}

Somewhat surprising, or maybe not, the paper was not easily
accepted. The first  referee simply rejected it as  (i) too
speculative and (ii) irrelevant because GRBs are galactic. The
second one accepted it reluctantly stating: {\it ``It three well
respected well known scientists what to make fools of themselves who
am I to stop them.}  I was very happy to be referred to as a
``respectable and well known scientist" but I suspected that I
gained this title because of my co-authors. I was even happier that
the paper was accepted. In retrospect given the ``anti-cosmological"
atmosphere at that time we should have been grateful to have had
such a broad minded referee.

\subsection{From Matter to Light – Internal Shocks}

The most urgent question was how to convert the relativistic
baryonic outflow to radiation. The solution proposed in NPP92 was
internal shocks: {\it ``These ejecta should, through collisions with
one another at large radii and low optical depth, give a non
Planckian spectrum by various nonthermal mechanisms. For instance
given the strong magnetic fields, synchrotron processes might
naturally produce the observed power-law spectrum."}. This was of
course just a short exposition of the idea and Bohdan continues to
elaborate on it later in 1993 together with Ghuohong Xu
(Paczy{\'n}ski \& Xu 1993). The basic idea is that the source is
irregular and it emits a wind with a variable Lorentz factor. An
idealized picture of such a wind is a situation when the outflow is
in the form of shells that moves with different Lorentz factors.
Faster shells collide with slower ones. Paczy{\'n}ski \& Xu (1993)
consider within this context proton proton collisions that produce
pions that decay producing Gamma-Rays. However this won't be very
efficient. Later on Rees \& M{\'e}sz{\'a}ros (1994)suggested the
picture accepted now according to which the collisionless shocks
that form accelerate particles to very high energies and possibly
also generate strong magnetic fields. These shocks are the source of
the observed prompt emission. The schematic picture is depicted as a
central part of Fig. 1 and in Fig. 4.  It is central now to our
current model. However at the time the simpler external shocks
picture proposed a few years earlier by M{\'e}sz{\'a}ros and Rees
(1992)  - namely interaction of the outflow with the surrounding
matter- was considered as the source of the prompt emission. Only in
1997  we (Sari \& Piran, 1997) have shown that external shocks
cannot produced the observed variable  light curves and as of today
internal shocks are the only known viable way to convert the kinetic
energy of the outflow to radiation and produce a highly variable
light curve!

\begin{figure}[!ht]
\plotfiddle{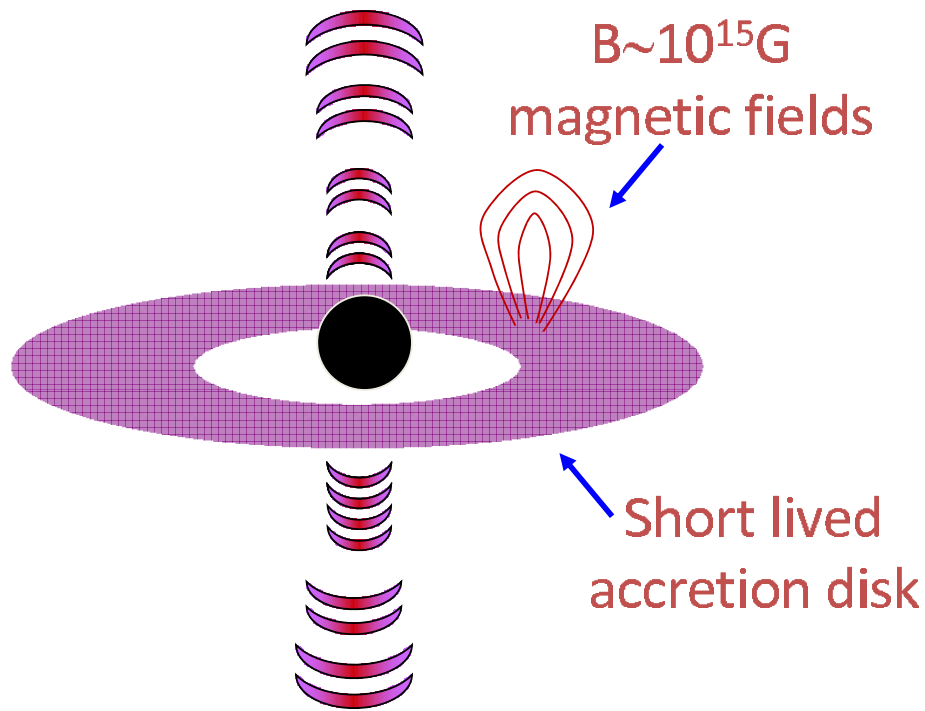}{10cm}{-90}{70}{70}{-310}{350} \caption{Internal
shocks are produced from a variably flow (schematically depicted as
shells). The faster shells catch up with the slower ones and produce
the shocks. The system is, most likely, powered by an accretion disk
and $10^{15}$ Gauss field could arise in this disk. }
\end{figure}

It should be stressed that the same ideas hold for Poynting flux
dominated outflow, a possibility that is considered seriously now as
an alternative to the baryonic outflow. The kinematic arguments that
require internal process within the outflow are applicable
regardless of the exact nature of the outflow. The main difference
of course is that in the Poynting flux case  instabilities such as
reconnection replace  the collisionless shocks. But the basic
feature that external interaction cannot produce the variable light
curves remains.

The idea of internal shocks is linked directly to another important
issue. In internal shocks the temporal structure of the burst is
determined by the inner engine. The duration is determined by the
time that the source is active while the dynamical time scale of the
source dictates the short term fluctuations time scale seen in the
individual pulses.  The need of a prolonged activity of the inner
has pointed out towards accretion onto a newborn black hole as the
maincandidate for the activity

\subsection{From Matter to Light –  Afterglow}

Clearly, not all the kinetic energy can be dissipated in  internal
shocks. Only the relative kinetic energy can be dissipated there.
The bulk center of mass kinetic  will remain. This leads immediately
to the idea of an afterglow: {\it ``The ejecta should much later
also produce something similar to a supernova remnant."} (NPP92).
With James Rhodas,  (Paczy{\'n}ski \& Rhodas, 1993) Bohdan continues
to elaborate on this idea. They use the analogy with supernova to
estimate that GRBs will be followed by a radio transients. More
detailed models focusing on higher energies were suggested later by
numerous authors, (M{\'e}sz{\'a}ros \& Rees, 1997; Wijers, Rees \&
M{\'e}sz{\'a}ros, 1997; Waxman, 1997;  Sari, Piran  \&  Narayan,
1998). Still this was the first suggestion that GRBs will be
followed by a long duration afterglow.

The afterglow prediction  was verified in 1997 with the discovery of
X-ray afterglow by BeppoSAX (Costa et al., 1997). This discovery was
followed by detection of optical (van Paradijs et al., 1997) and
radio (Frail et al., 1997) afterglows as well.

\section{GRB progenitors - mergers and hypernovae}

The question what makes GRBs is of course the central one for the
whole problem. Once it was realized that the bursts are cosmological
and the energy budget was set to be around $10^{51}-10^{52}$ergs it
was clear that the sources involve a compact object. Most likely the
formation of the object and the release of its binding energy, and
if not that at least a significant catastrophic event. The neutron
star merger model, which is currently the most likely to work for
short GRBs, was proposed already in the eighties (Paczy{\'n}ski,
1986; Eichler et al., 1999). For a while it was thought that this is
the model for all GRBs.

However, when BeppoSAX begun detecting X-ray afterglows and host
galaxies were revealed, it turned out that GRBs (actually long GRBs,
as BeppoSAX detected afterglow only from long GRBs) are located
within small irregular star forming galaxies. Now we know that
within these galaxies GRBs are located in the highest star forming
regions and that the rate of GRBs is roughly proportional to the
square of the rate of star formation (Fruchter et al., 2006). Bohdan
did not need much data to make a far reaching conclusion. At the
fall of 1997 he concluded on the basis of just three well localized
GRBs, 970228, 0970508 and 980828 that GRBs arise in the vicinity of
star forming regions (Paczy{\'n}ski, 1998). He concludes that the
merger model must be abandoned and   GRBs must be linked to death of
massive stars. {\it ``There is tentative evidence that the GRBs
970228, 970508, and 970828 were close to star-forming regions. If
this case is strengthened with future afterglows, then the popular
model in which GRBs are caused by merging neutron stars will have to
be abandoned, and a model linking GRBs to cataclysmic deaths of
massive stars will be favored."}

In a typical manner Bohdan ignored the theoretical prejudice against
this idea. At the time baryonic contamination was a major concern.
Within the fireball model discussed above it was clear that a
significant baryonic load will result in a non-relativistic outflow
which won't be able to drive a GRB. Collapsing stars have large
envelopes that could be a strong source of contamination. However,
observations are more important than theory and if the bursts are
near star forming regions they must involve stellar death. Bohdan
outlines a model based on Woosley's (1993) ``failed supernova" on
accretion and Blandford-Znajek mechanism (1976) and suggests that
GRBs operate like microquasars.

Once again Bohdan's ability to grasp the basic point from a dismal
amount of data leads to a great success. A few month later Galama et
al., (1998) discovered that a very powerful  type Ic SN 1998bw is
associated with GRB 980425! At the same time MacFadyen \& Woosley
(1999) demonstrate that a sufficiently powerful jet can punch a hole
in a stellar atmosphere, (provided that the later is sufficiently
small as would be the case if the Hydrogen envelope has escaped). As
more and more SN like bumps were discovered on GRB afterglow light
curves (Bloom et al., 1999) the model gained credibility. It was
eventually proven with GBR 030339/SN 2003dh where an 98bw like SN
spectrum arose just as expected from the GRB optical light curve
(Staneck et al., 2003; Hjorth, J., et al. 2003).

In the process Bohdan coined the name Hypernovae (Paczy{\'n}ski,
1998).  Bohdan anticipated that GRB associated collapse events are
more powerful than a regular supernovae and hence they should be
given a name indicating that. Hyper is clearly more powerful than
Super. It turned out that the first GRB associated supernova SN
1998bw was indeed much more powerful than average and indeed GRB
associated supernovae do show higher velocity  ejecta and are more
powerful than other type Ic SNe.

\section{Some ideas for the future: Neutrinos, GRB remnants and optical flashes
preceding GRBs}

Bohdan has left many ideas to be tested in the future.

\subsection{GRB neutrinos}
Already in 1994 Bohdan noticed (Paczy{\'n}ski \& Xu, 1993) that
internal shocks within GRBs can produce high energy neutrinos. In
this specific model proton-proton collisions between protons from
different shells produce pions,which in turn produce the observed
gamma-rays as well as 30 GeV neutrinos. In fact, more energy is
released in this case in neutrinos than in gamma-rays.

Proton-proton collisions are not very efficient and they can produce
only a weak GRB signal. To overcome this problem Paczy{\'n}ski and
Xu proposed that the bursts are very narrow beamed. As far as we
understand GRBs today both ideas are invalid. GRBs are beamed but
the jets are much wider. Proton-proton collisions are indeed not
efficient enough to produce the observed prompt emission. However,
what we do take from this paper today is first the basic concept of
internal shocks within the outflows and second the idea that GRBs
are prime candidates for being sources of high energy neutrinos - an
idea that is generally accepted today (see e.g. Achterberg et
al.(2007) for a description of a recent search using AMANDA).

\subsection{GRB remnants}
Several times we have worked in parallel on related idea. The last
one was in the early 2000. It was realized at that time that GRBs
are beamed (Rhodas, 1999; Sari et al., 1999; Kulkarni et al., 1999).
It was also realized that the afterglow, in which the Lorentz factor
is much lower is essentially less beamed. One expects therefore,
orphan afterglows. Afterglows whose GRB prompt emission does not
points towards us. An intriguing question was how to search for such
orphan afterglows. The late radio phase is naturally the longest
one. But this phase is the weakest. Additionally, if one waits too
long the afterglow becomes spherical and indistinguishable from a
regular Supernova remnant.

\begin{figure}[!ht]
\plotfiddle{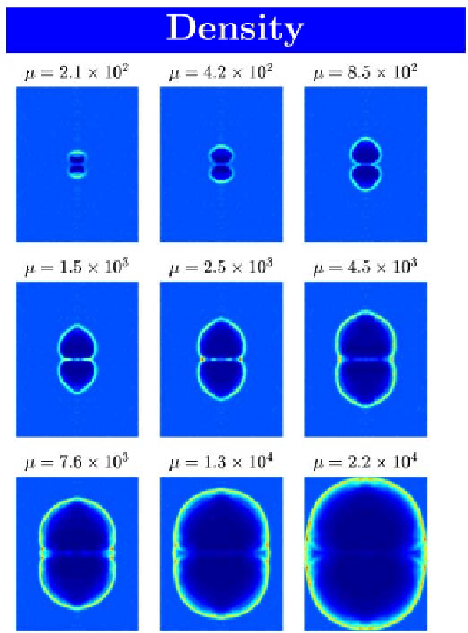}{10cm}{-90}{90}{90}{-350}{370}
\caption{Density profile of a GRB remnant as a function of $ \mu
\equiv Mc^2 /E_0$, where M is the accumulatd mass and $E_0$ is the
initial energy (from Ayal \& Piran, 2001). }
\end{figure}

However, for a period of 5000  years it is possible to see within
the GRB remnant a bipolar structure that is induced by the original
nonspherical GRB jets (Ayal \& Piran, 2001). This will enable us to
distinguish GRB remnants from SNRs.  Bohdan (Paczy{\'n}ski, 2001a)
build on these ideas and estimated that at any time there should be
several dozen relatively nearby GRB radio remnants which can be
resolved with VLBA as being bipolar rather than spherical. He
suggested to combine this idea with other methods to detect imprints
of GRBs such as the effect of the original gamma-rays on the
interstellar medium (Draine, 2000) and perform a systematic search
for GRB remnants. Such a search was not done yet. It is something
that Bohdan has left for the future.

\subsection{Optical Flashes}

In recent years Bohdan was fascinated with the transient Universe
(Paczy{\'n}ski, 2001). This was not surprising. After all both
microlensing events and GRBs to whom he devoted his research in
recent years are transient phenomenon. He kept constantly looking
for new possible ``flashes" that will shine in the night sky. Along
this line Bohdan noticed  an exciting idea of Beloborodov (2002)
that the prompt emission  can cause a cascade of pairs at a distance
of $\sim 10^{18}$cm from the origin. The idea is simple. Some
gamma-rays will interact with the surrounding matter and will be
reflected backwards. Each photon that is reflected backwards will
certainly interact with the outgoing radiation and produce a pair.
The produced pairs will increase the number of back scattered
photons and will accelerate the process causing a possible runaway.
Bohdan suggested that with the right conditions (Kumar \&
Panaiteschu, 2004) a cloud of pairs will form and this cloud will be
opaque to gamma-rays. Until this cloud is cleared away only lower
energy photons will be seen.  Bohdan suggested, therefore, that an
optical flash will appear in such a case and that this flash will
precede the gamma-rays (Paczy{\'n}ski, 2001). He was worried,
however, that such flashes might be missed, as they appear before
the GRB trigger, and he searched for method to detect them.

\section{Epilogue}

 Bohdan was dominant in convincing the community that GRBs are
cosmological. However, he also had far reaching contributions that
shaped our current understanding of how GRBs operate. The scope of
these contribution is best realized by presenting once again Fig. 1,
that describes the basic theory of GRBs, but now with Bohdan's
contributions superimposed on it. His ideas on hypernova, links with
star formation, internal shocks, afterglow that looks like a SNR,
and above all the cosmological origin of GRBs all paved the ways to
the present ``standard" model.

\begin{figure}[!ht]
\plotfiddle{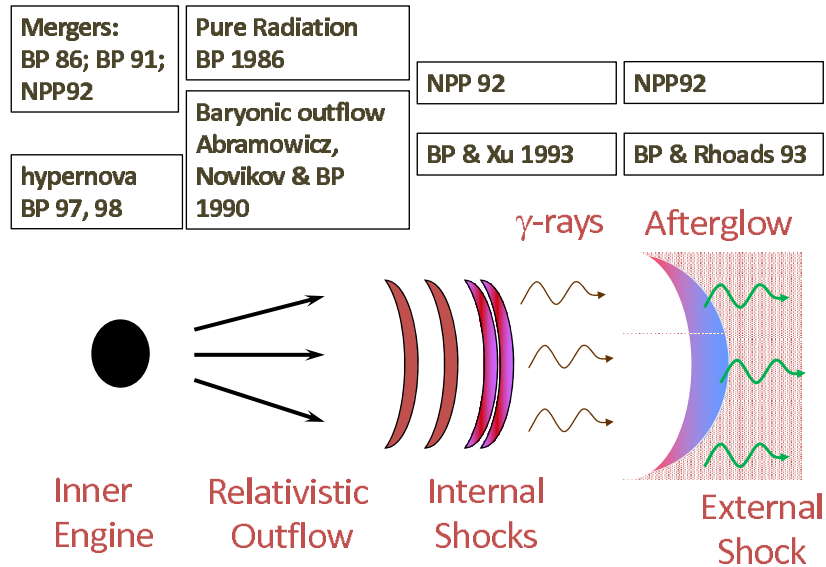}{8cm}{-90}{50}{50}{-200}{290}
\caption {Bohdan's major contributions to the current GRB model.}
\end{figure}

The fast response satellite {\it Swift} that was launched three
years ago have changed to a large extend the simple picture that we
had before. It turns out that, like in other cases in Astronomy, the
afterglow picture is more complicated than what was originally
thought. In particular the early X-ray light curve, as seen by {\it
Swift} is rather different from the previous expectations, showing
unexpected rapid decline that is followed by a shallow phase before
joining the more familiar light curve at about $10^4$sec (Nousek et
al., 2006). At present it is not clear what are the processes that
control this light curve. One widely discussed possibility is that
energy is added to the blast wave during the shallow phase. It is
interesting to note that already in 1993 Bohdan  discussed the
possibility that a significant amount of energy can come out from
the inner engine in a low Lorentz factor material (Paczy{\'n}ski \&
Xu, 1993). ``The slower material with a low Lorentz factor will be
gradually added to the blast wave". This kind of ``energy injection"
is in fact the leading interpretation today of this phase (see e.g.
Zhang et al., 2006; Granot, Konigl \&  Piran, 2006).

Another interesting issue is the fact that {\it Swift} has observed
bursts that are further out than what is expected if the bursts
simply follow the SFR (Natarajan et al., 2005,, Daigne et al., 2006;
Guetta \& Piran, 2007). There are more distance bursts than the
simple model suggests. This result is consistent with the fact that
in space GRBs are more concentrated in high SFR regions, as if they
follow a higher power of the SFR. It could be that the relevant
factor is low metalicity (Fynbo, 2003) but the possibility of
evolution of the luminosity function or another unexpected factor
cannot be ignored.

Other long standing open questions have  existed before {\it Swift}
and are still with us: What is  the exact working of the inner
engine, what is the nature of the relativistic outflow and how do
collisionless shock accelerate particles and generate magnetic
fields? It is illuminating to read what were Bohdan's thoughts about
all that: {\it ``It is not likely that the concept of a GRB as a
microquasar powered by the Blandford \& Znajek (1977) mechanism can
be proven or disproven on purely theoretical grounds. It is useful
to realize, that while we have plenty of sound evidence that Type II
supernovae explode as a result of some 'bounce', or whatever process
following the formation of a hot neutron star, there is no generally
accepted physical process which would be efficient enough to make
this happen. The theoretical problem with the SN II explosions
persists in spite of 2 or 3 decades of intense effort by a large
number researchers. The problem is vastly worse with the GRBs as
they are $10^4-10^5$ times less common than supernovae. This might
imply that a very special set of circumstances is necessary to
generate the suitable energetic explosion". }

Bohdan was probably right. It will take time, a lot of detailed
observations and ingenious theoretical insight to figure out all
those details. However, regardless of the different variants of the
current model and the current observational puzzles I am sure that
Bohdan's basic ideas on how GRBs work will be with us forever  to
stay.


\acknowledgements 
This work was partially supported by an ISF center for Excellence in
High Energy Astrophysics. 


\end{document}